\documentstyle[aps,epsfig,floats]{revtex}
\newcommand{\bm}[1]{\mbox{\boldmath $#1$}}
\newcommand{\be}{\begin{equation}}
\newcommand{\ee}{\end{equation}}
\newcommand{\bea}{\begin{eqnarray}}
\newcommand{\eea}{\end{eqnarray}}
\newcommand{\st}{{\scriptscriptstyle T}}
\newcommand{\xbj}{x_{\scriptscriptstyle B}}
\newcommand{\zh}{z_h}
\def\slash{\rlap{/}}

\begin{document}
\tighten
\thispagestyle{empty}
\title{
\begin{flushright}
\begin{minipage}{3 cm}
\small
hep-ph/9711485\\ 
NIKHEF 97-049\\
VUTH 97-20
\end{minipage}
\end{flushright}
Time-reversal odd distribution functions in leptoproduction } 
\author {D. Boer$^{a}$ and P.J. Mulders$^{a,b}$\\  
\mbox{}\\
{\it $^{a}$National Institute for Nuclear Physics and High-Energy Physics}\\
{\it P.O. Box 41882, NL-1009 DB Amsterdam, the Netherlands}\\
\mbox{}\\
{\it $^{b}$Department of Physics and Astronomy, Free University of Amsterdam}\\
{\it De Boelelaan 1081, NL-1081 HV Amsterdam, the Netherlands}}

\maketitle

\begin{abstract}
We consider the various asymmetries, notably single spin asymmetries, that 
appear in leptoproduction as a consequence of the presence of 
time-reversal odd distribution functions. 
This could facilitate experimental searches for time-reversal odd phenomena. 
\end{abstract}

  
In this paper we study the effects of the possible presence of
time-reversal (T) odd distribution functions in leptoproduction.
We limit ourselves to the production of hadrons in the current
fragmentation region, for which we assume that the
cross section factorizes into a product of a distribution function 
and a fragmentation function.
Including the effects of transverse momenta, the cross section
is assumed to factorize into a convolution of distribution and 
fragmentation functions which not only depends on the lightcone momentum
fractions of quark and hadron, but also on the transverse momentum of
quark with respect to hadron or vice versa~\cite{mt96}.

Starting with the expressions of the soft parts in hard scattering
processes as quark-quark lightfront correlation functions, 
i.e.\ matrix elements of nonlocal combinations of quark fields, one
can analyze the various possible distribution and
fragmentation functions. Constraints arise from Lorentz
invariance, hermiticity, parity invariance and time-reversal
invariance. The latter, however, cannot be used as a constraint
on fragmentation functions, because the produced hadron can
interact with the debris of the fragmenting quark, a well-known
phenomenon in any decay process~\cite{Gas66}.
This allows so-called T-odd quantities, although it is hard to
say something about their magnitude. In Ref.~\cite{jjt97} it was even
conjectured that final state interaction phases average to zero
for single hadron production after summation over unobserved final states.

Without considering transverse momenta of quarks, the T-odd effects are 
higher twist, appearing at order $1/Q$~\cite{ji94}. 
Including transverse momenta of quarks, there are leading order effects.
One can have fragmentation of transversely polarized quarks
into unpolarized or spin zero hadrons or production of transversely
polarizated hadrons in the fragmentation of unpolarized quarks~\cite{mt96}.  
For the distribution functions, it has been conjectured that T-odd
quantities also might appear
without violating time-reversal invariance~\cite{s90,abm95,alm96,adm96}.
This might be due to soft initial state interactions or, as suggested
recently\cite{adm96}, be a consequence of chiral symmetry breaking.
Within QCD a possible description of the effects may come from gluonic
poles\cite{btm97}.

It is convenient to use the hadron momenta in the process 
$\ell H \rightarrow \ell^\prime h X$ to define two lightlike 
vectors $n_+$ and $n_-$, satisfying $n_+\cdot n_-$ = 1. 
These vectors then define the lightcone components of a vector as $
a^\pm \equiv a\cdot n_\mp$.  Up to mass terms the momentum $P$ of the target
hadron ($H$) is along $n_+$, that of the outgoing hadron along $n_-$.
We assume here that we are discussing current fragmentation, for which
one requires $P\cdot P_h \sim Q^2$, where $q^2 = - Q^2$ is the
momentum transfer squared. In leading order in $1/Q$ the process
factorizes into a product of two soft parts. For the description of
the quark content of the target the following quantity (given in
the lightcone gauge $A^+ = 0$) is relevant,
\be
\Phi(x,\bm p_\st) = \left. \int \frac{d\xi^-d^2\bm \xi_\st}{(2\pi)^3} 
\ e^{ip\cdot \xi} \,\langle P,S \vert \overline \psi (0) 
\psi(\xi) \vert P,S \rangle \right|_{\xi^+ = 0},
\ee
depending on the lightcone fraction of the quark momentum, $x = p^+/P^+$ 
and the transverse momentum component $\bm p_\st$.
Using Lorentz invariance, hermiticity, and parity invariance one finds that
the Dirac projections that will appear in a calculation up to 
leading order in $1/Q$ can be expressed in a number of distribution
functions
\bea
\Phi(x,\bm p_\st) &=& 
\frac{1}{2}\,\Biggl\{
f_1\,\slash \slash n_+
+ f_{1T}^\perp\, \frac{\epsilon_{\mu \nu \rho \sigma}
\gamma^\mu n_+^\nu p_\st^\rho S_\st^\sigma}{M}
+g_{1s}\, \gamma_5\slash n_+
\nonumber \\ & & \qquad
+h_{1T}\,i\sigma_{\mu\nu}\gamma_5 n_+^\mu S_\st^\nu 
+h_{1s}^\perp\,\frac{i\sigma_{\mu\nu}\gamma_5 
n_+^\mu p_\st^\nu}{M} 
+ h_1^\perp\,\frac{\sigma_{\mu \nu} p_\st^\mu n_+^\nu}{M}
\Biggr\},
\eea
with arguments $f_1$ = $f_1(x,\bm p_\st^2)$ etc. The quantity $g_{1s}$ 
(and similarly $h_{1s}^\perp$) is shorthand for 
\be 
g_{1s}(x,\bm p_\st) = 
\lambda\,g_{1L}(x,\bm p_\st^2) + \frac{\bm p_\st\cdot \bm S_\st}{M}
\,g_{1T}(x,\bm p_\st^2),
\ee
with $M$ the mass, $\lambda = M\,S^+/P^+$ the lightcone helicity, 
and $\bm S_\st$ the transverse spin of the target hadron.
Note that the difference with the analysis in Ref.~\cite{mt96}, in 
which the time-reversal constraint has been imposed, is the appearance
of the functions $f_{1T}^\perp$ and $h_1^\perp$. The function $f_{1T}^\perp$
is interpreted as the unpolarized quark distribution in a transversely
polarized nucleon, while $h_1^\perp$ is interpreted as the quark transverse 
spin distribution in an unpolarized hadron. 
We have used and followed the naming convention of Ref.~\cite{mt96}.
The function $f_{1T}^\perp$ is proportional to the function 
$\Delta^N f$ used in 
Refs~\cite{abm95,adm96}. In this paper we simply want to investigate
where the functions $f_{1T}^\perp$ and $h_1^\perp$ show up in leptoproduction.
We do not discuss the possible mechanisms leading to them, but point out
in which observables their existence can be checked experimentally.

The computation of the leading order leptoproduction cross sections 
requires in addition to the quark distribution functions, 
also fragmentation functions, contained in a soft part 
which (in the lightcone gauge $A^- = 0$) is of the form
\be
\Delta(z,\bm k_\st) = 
\left. \sum_X \int \frac{d\xi^+d^2\bm\xi_\st}{2z\,(2\pi)^3} \ 
e^{ik\cdot \xi} \,\langle 0 \vert \psi (\xi) \vert X;P_h,S_h\rangle
\langle X;P_h,S_h\vert \overline \psi(0) \vert 0 \rangle 
\right|_{\xi^- = 0},
\ee
where $z = P_h^-/k^-$ is the lightcone fraction of the produced hadron
and $\bm k_\st$ is the quark transverse momentum with respect to the 
produced hadron, which implies a transverse momentum $\bm k_\st^\prime 
= -z\,\bm k_\st$ of the produced hadron with respect to the fragmenting 
quark. At leading order the following expansion in fragmentation functions
can be written,
\bea
\Delta(z,\bm k_\st) & = &
\frac{1}{2} \Biggl\{
D_1\,\slash n_-
+ D_{1T}^\perp\, \frac{\epsilon_{\mu \nu \rho \sigma}
\gamma^\mu n_-^\nu k_\st^\rho S_{hT}^\sigma}{M_h}
+ G_{1s}\,\gamma_5 \slash n_- 
\nonumber \\ & & \qquad 
+ H_{1T}\,i\sigma_{\mu\nu}\gamma_5 \,n_-^\mu S_{h\st}^\nu 
+ H_{1s}^\perp\,\frac{i\sigma_{\mu\nu}\gamma_5\,n_-^\mu k_\st^\nu}{M_h} 
+ H_1^\perp\,\frac{\sigma_{\mu \nu} k_\st^\mu n_-^\nu}{M_h}
\Biggr\},
\eea
with arguments $D_1$ = $D_1(z,z^2\bm k_\st^2)$ etc. The quantity
$G_{1s}$ (and similarly $H_{1s}^\perp$) is shorthand for 
\be 
G_{1s}(z,-z\bm k_\st) = 
\lambda_h\,G_{1L}(z,z^2\bm k_\st^2) + \frac{\bm k_\st\cdot \bm S_{h\st}}{M_h}
\,G_{1T}(z,z^2\bm k_\st^2),
\ee
with $M_h$ the mass, $\lambda_h = M_h\,S_h^-/P_h^-$ the lightcone helicity, 
and $\bm S_{h\st}$ the transverse spin of the produced hadron.
The functions $D_{1T}^\perp$ and $H_1^\perp$ are the T-odd ones
in the fragmentation part. Although written down for spin-1/2 hadrons,
all results will include also target hadrons and produced hadrons with
spin zero (putting $S$ = 0 or $S_h$ = 0).

\begin{figure}[t]
\begin{center}
\leavevmode \epsfxsize=10cm \epsfbox{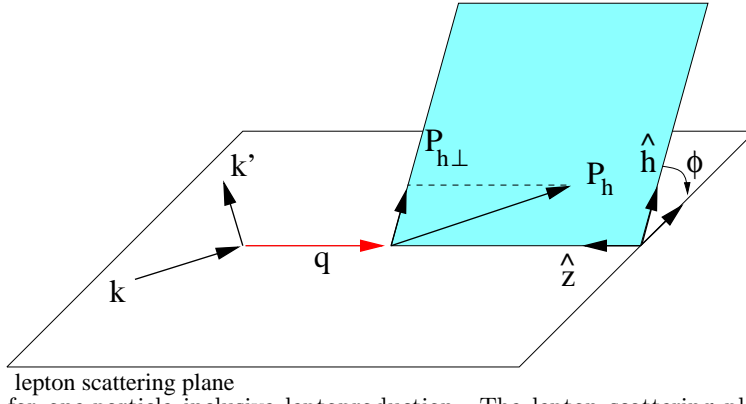}
\caption{\label{fig1} Kinematics for one-particle inclusive leptoproduction.
The lepton scattering plane is determined by the momenta
$k$, $k^\prime$ and $P$.}
\end{center}
\end{figure}

The T-odd functions appear in pairs in the unpolarized 
leptoproduction cross section or in double spin asymmetries and they
appear singly in single spin asymmetries. 
The hadron tensor in leading order in $1/Q$ (thus also neglecting all 
mass corrections) is given by
\bea
2M\,{\cal W}_{\mu \nu}(q,P,P_h) =
\int d^2\bm p_\st\,d^2\bm k_\st 
\,\delta^2(\bm p_\st + \bm q_\st - \bm k_\st)
\,\frac{1}{4}\,\mbox{Tr} \left( \Phi(\xbj, p_\st) \gamma_\mu 
\Delta(\zh, k_\st) \gamma_\nu \right) 
+ \left(\begin{array}{c} q\leftrightarrow -q \\ \mu \leftrightarrow \nu
\end{array} \right),
\eea
where $\xbj = Q^2/2P\cdot q$ and $\zh = P\cdot P_h/P\cdot q$. The momentum 
$q_\st^\mu$ is the transverse momentum of the exchanged photon in the
frame where $P$ and $P_h$ do not have transverse momenta, which is
proportional to the transverse component of the produced hadron, 
$P_{h\perp}^\mu$, in the frame where $P$ and $q$ have no transverse 
components. In general we will indicate
transverse momenta in the first frame with a subscript $T$ 
(thus $P_\st = 0$ and $P_{h\st} = 0$) 
and those in the second frame with a subscript $\perp$
(thus $P_\perp = 0$ and $q^{}_\perp = 0$). The kinematics for 
one-particle inclusive leptoproduction in the second frame have been
shown in Fig.~\ref{fig1}.
One has
\be
q_\st^\mu \ = \ \left(g^{\mu\nu} - n_+^{\{\mu} n_-^{\nu\}}\right) q_\nu
\ = \ q^\mu + \xbj\,P^\mu - \frac{P_h^\mu}{\zh} \ =\ -\frac{P_{h\perp}}{\zh}
\equiv -Q_\st\,\hat h^\mu.
\ee
It is convenient to 
introduce the tensors $g_\perp^{\mu\nu}$ and $\epsilon_\perp^{\mu\nu}$ given by
\bea
&&g_\perp^{\mu\nu} = g^{\mu\nu} - \frac{q^\mu q^\nu}{q^2} 
+ \frac{\tilde P^\mu \tilde P^\nu}{\tilde P^2},
\\
&&\epsilon_\perp^{\mu\nu} = \frac{\epsilon^{\mu\nu\rho\sigma}P_\rho q_\sigma}
{P\cdot q},
\eea
where $\tilde P^\mu$ = $P^\mu - (P\cdot q/q^2)\,q^\mu$. The tensors
act in the transverse space orthogonal to $P$ and $q$. 
If $Q_\st \ll Q$ one has $(p_\st)_\perp \approx p_\st$, 
$(k_\st)_\perp \approx k_\st$, $(S_\st)_\perp \approx S_\st$, 
and $(S_{h\st})_\perp \approx S_{h\st}$. Azimuthal angles
will be defined in this space with respect to the lepton
scattering plane (see Fig.~\ref{fig1}), e.g.\ $\phi_h^\ell = \phi_h
-\phi^\ell$ is the angle between hadron production plane (defined by
$P_h$ and $q$) and the lepton scattering plane.

We will next discuss the explicit results for the cross sections. At 
leading order they are obtained from the contraction of the lepton
tensor with the hadron tensor ${\cal W}^{\mu\nu}$.
We will in general consider cross sections integrated over the transverse
momentum of the produced hadron (i.e.\ over $\bm q_\st$) and depending on
the weight denote them by
\be
\langle W\rangle_{ABC}
= \int d\phi^\ell\,d^2\bm q_\st
\ W\,\frac{d\sigma_{ABC}^{[\vec e\vec H\rightarrow e\vec hX]}}
{d\xbj\,dy\,dz_h\,d\phi^\ell\,d^2\bm q_\st},
\label{asym}
\ee
where $W$ = $W(Q_\st,\phi_h^\ell,\phi_S^\ell,\phi_{S_h}^\ell)$.
In order to see in a glance which polarizations 
are involved, we have added the subscripts ABC for polarizations of
lepton, target hadron and produced hadron, respectively. 
We use O for unpolarized, 
L for longitudinally polarized ($\lambda \ne 0$) 
and T for transversely polarized ($\vert \bm S_\st\vert \ne 0$) particles.

Starting (as a reference) with the cross section for {\em unpolarized 
leptons scattering off an unpolarized hadron producing a spin zero particle 
or summing over spin in the final state}, one finds
\be
\left<1\right>_{OOO}
\ =\ \frac{4\pi \alpha^2\,s}{Q^4}
\,\left\lgroup 1-y+\frac{y^2}{2}\right\rgroup \sum_{a,\bar a} e_a^2
\,\xbj\,f_1^a(\xbj) D_1^a(\zh).
\ee
The above is the well-known unpolarized result containing a sum over 
flavors of quarks and antiquarks with in each
term the product of the unpolarized distribution function $f_1^a$ (quarks
$a$ in hadron $H$) and the unpolarized quark fragmentation function $D_1^a$
(quark $a$ fragmenting into hadron $h$). 
Only considering T-even distribution functions, this is the only nonvanishing
averaged unpolarized cross section at leading order. At subleading ($1/Q$)
order one has a nonvanishing $\cos \phi_h^\ell$-asymmetry originating from
kinematical~\cite{c78} and dynamical~\cite{lm94} effects, while a
$\cos 2\phi_h^\ell$-asymmetry only appears at order $1/Q^2$~\cite{b80,OABD97}.
Allowing for T-odd functions, however, the following weighted
cross section projects out a {\em leading} azimuthal 
$\cos 2\phi_h^\ell$-asymmetry,
\be
\left< \frac{Q_\st^2}{MM_h} \,\cos(2\phi_h^\ell)\right>_{OOO}
\ =\ \frac{16\pi \alpha^2\,s}{Q^4}
\,\left\lgroup 1-y\right\rgroup\sum_{a,\bar a} e_a^2
\,\xbj\,{h_{1}^{\perp(1)a}}(\xbj) H_1^{\perp (1)a}(\zh).
\ee
The weighted cross section involves $\bm p^2_\st$-moments 
of the distribution and fragmentation functions $h_1^\perp$ and
$H_1^\perp$, defined as
\bea
h_1^{\perp (n)}(x) & \equiv & \int d^2\bm p_\st 
\,\left(\frac{\bm p_\st^2}{2M^2}\right)^n
\,h_1^\perp(x,\bm p_\st),
\\
H_1^{\perp (n)}(z) & \equiv & z^2\int d^2\bm k_\st 
\,\left(\frac{\bm k_\st^2}{2M_h^2}\right)^n
\,H_1^\perp(z,-z\bm k_\st).
\eea
While the $\bm k_\st$-dependent function are lightfront correlation
functions (i.e.\ $\xi^+$ = 0), the 
integrated functions and $\bm k_\st^2$-moments 
are lightcone correlation functions
(i.e.\ $\xi^+ = \xi_\st$ = 0 in the matrix elements) for which we expect
factorization to remain valid, although this has not yet been proven.
The above two cases are summarized in Table~\ref{tabel1}.
\begin{table}[htb]
\caption{\label{tabel1}\protect
Azimuthal asymmetries $\left< W \right>_{ABC}$ (see Eq.~\ref{asym}) 
for the case of fully unpolarized leptoproduction.
The last column indicates the time-reversal behavior
of the distribution and fragmentation function, respectively (e = even,
o = odd).}
\begin{tabular}{cclr}
\\[-2 mm]
$ABC$ & $W$ 
& $\left<W\right>_{ABC} \cdot \left[4\pi\,\alpha^2\,s/Q^4\right]^{-1} $
& T 
\\[2mm] \hline 
\\[-2 mm]
OOO & 
1 & 
$\left( 1-y + \frac{1}{2}\,y^2\right) \sum_{a,\bar a} e_a^2
\,\xbj\,f_1^a(\xbj) D_1^a(\zh)$
& ee 
\\[1mm]
OOO & 
$(Q_\st^2/4MM_h)\,\cos(2\phi_h^\ell)$ & 
$(1-y)\sum_{a,\bar a} e_a^2
\,\xbj\,{h_{1}^{\perp(1)a}}(\xbj) H_1^{\perp (1)a}(\zh)$
& oo
\end{tabular}
\end{table}
We note that a similar $\cos 2\phi$ asymmetry involving the azimuthal
angle of two hadrons in opposite jets appears in
electron-positron annihilation~\cite{bjm97}. In that case only
T-odd fragmentation functions $H_1^{\perp (1)}$ and 
$\overline H_1^{\perp (1)}$ are involved. A similar asymmetry in
Drell-Yan would involve only T-odd distribution functions.

Next, we consider leading order single spin asymmetries, which
we separate in single spin asymmetries for lepton, target hadron
and produced hadron, respectively. 
There are no leading order lepton spin asymmetries. The one lepton spin
asymmetry that is possible in one-particle inclusive leptoproduction
is a $\sin \phi_h^\ell$ asymmetry. It is, however, subleading, i.e.\ order
$1/Q$ (see Ref.~\cite{lm95}). 

There are four leading order single spin asymmetries involving the spin 
of the target hadron, given in Table~\ref{tabel2}. The first three
involve T-even distribution functions. The fourth one involves a
T-odd distribution function.
\begin{table}[htb]
\caption{\label{tabel2}\protect
Leading order single spin asymmetries for the case 
of leptoproduction into unpolarized final states.}
\begin{tabular}{cllc}
\\[-2 mm]
$ABC$ & $W$ 
& $\left<W\right>_{ABC} \cdot \left[4\pi\,\alpha^2\,s/Q^4\right]^{-1} $
& T 
\\[2mm] \hline 
\\[-2 mm]
OLO &
$(Q_\st^2/4MM_h)\,\sin(2\phi_h^\ell)$  & 
$-\lambda\,(1-y) \sum_{a,\bar a} e_a^2
\,\xbj\,h_{1L}^{\perp (1) a}(\xbj) H_1^{\perp (1) a}(\zh)$
& eo
\\[1mm]
OTO &
$(Q_\st/M_h)\,\sin(\phi_h^\ell+\phi_S^\ell)$  & 
$\vert \bm S_\st\vert\,(1-y) \sum_{a,\bar a} e_a^2
\,\xbj\,h_{1}^{a}(\xbj) H_1^{\perp (1) a}(\zh)$
& eo
\\[1mm]
OTO &
$(Q_\st^3/6M^2M_h)\,\sin(3\phi_h^\ell-\phi_S^\ell)$ & 
$\vert \bm S_\st\vert\,(1-y) \sum_{a,\bar a} e_a^2
\,\xbj\,h_{1T}^{\perp(2)a}(\xbj) H_1^{\perp (1) a}(\zh)$
& eo 
\\[2mm] \hline 
\\[-2 mm]
OTO &
$(Q_\st/M)\,\sin(\phi_h^\ell-\phi_{S}^\ell)$ & 
$\vert \bm S_{\st}\vert
\,\left( 1-y + \frac{1}{2}\,y^2\right) \sum_{a,\bar a} e_a^2
\,\xbj\,f_{1T}^{\perp (1) a}(\xbj) D_{1}^{a}(\zh)$
& oe
\end{tabular}
\end{table}

The first three asymmetries are T-odd in the fragmentation part, 
in particular they all feature the fragmentation function $H_1^{\perp (1)}$. 
The second of the three asymmetries was first discussed by 
Collins~\cite{col}. The existence of the other two was pointed out in 
Refs~\cite{ak95,tm95}. The third asymmetry in 
Table~\ref{tabel2} involves the second $\bm p_\st^2/2M^2$ moment of 
the function $h_{1T}^\perp$ and appears also as an OTO asymmetry
with slightly different azimuthal angle dependence.
For a detailed discussion of the asymmetries we refer to 
Refs~\cite{km96,km97}.  The fourth entry in Table~\ref{tabel2} is again
an OTO single spin asymmetry containing the T-odd distribution function 
$f_{1T}^{\perp (1)}$. It appears in
{\em scattering of unpolarized leptons off transversely polarized targets}.
This asymmetry is probably the easiest way to look for the function
$f_{1T}^\perp$ as it just requires searching for a correlation between
the azimuthal angles of the produced hadron and the target transverse spin,
e.g.\ in $ep^\uparrow \rightarrow e\pi X$ or $ep^\uparrow \rightarrow eK X$. 
This possibility was pointed out in Ref.~\cite{alm96} (measurement a).

Next we consider the single spin asymmetries related to the spin of the 
produced hadron.  
At leading order there are four single spin asymmetries, given in 
Table~\ref{tabel3}, three of which contain the T-odd distribution function
$h_1^{\perp (1)}$. 
\begin{table}[htb]
\caption{\label{tabel3}\protect
Leading order single spin asymmetries due to T-odd
distribution functions that require polarimetry.}
\begin{tabular}{cllc}
\\[-2 mm]
$ABC$ & $W$ 
& $\left<W\right>_{ABC} \cdot \left[4\pi\,\alpha^2\,s/Q^4\right]^{-1} $
& T 
\\[2mm] \hline 
\\[-2 mm]
OOT & 
$(Q_\st/M_h)\,\sin(\phi_h^\ell-\phi_{S_h}^\ell)$ & 
$-\vert \bm S_{h\st}\vert\,\left(1-y + \frac{1}{2}\,y^2\right) 
\sum_{a,\bar a} e_a^2
\,\xbj\,f_{1}^{a}(\xbj) D_{1T}^{\perp (1) a}(\zh)$
& eo
\\[2mm] \hline 
\\[-2 mm]
OOL &
$(Q_\st^2/4MM_h)\,\sin(2\phi_h^\ell)$  & 
$-\lambda_h\,(1-y) \sum_{a,\bar a} e_a^2
\,\xbj\,h_{1}^{\perp (1) a}(\xbj) H_{1L}^{\perp (1) a}(\zh)$
& oe
\\[1 mm]
OOT &
$(Q_\st/M)\,\sin(\phi_h^\ell+\phi_{S_h}^\ell)$  & 
$-\vert \bm S_{h\st}\vert\,(1-y) \sum_{a,\bar a} e_a^2
\,\xbj\,h_{1}^{\perp (1) a}(\xbj) H_1^{a}(\zh)$
& oe
\\[1 mm]
OOT &
$(Q_\st^3/6MM_h^2)\,\sin(3\phi_h^\ell-\phi_{S_h}^\ell)$ & 
$-\vert \bm S_{h\st}\vert\,(1-y) \sum_{a,\bar a} e_a^2
\,\xbj\,h_{1}^{\perp (1) a}(\xbj) H_{1T}^{\perp (2) a}(\zh)$
& oe
\end{tabular}
\end{table}

The latter three odd-even asymmetries appear in
{\em unpolarized lepton scattering off an unpolarized target hadron}.
They require polarimetry in the final state and are the direct counterparts 
of the three even-odd asymmetries in Table~\ref{tabel2} with the role of 
distribution and fragmentation functions being reversed.
These asymmetries can for instance be measured in
$ep \rightarrow e\Lambda^\uparrow X$, by determining the 
$\Lambda$ polarization and its orientation from the $p\pi^-$ final state.
At this point it may be good to reiterate the interpretation of the single 
spin asymmetries. In all cases a T-odd effect is needed in either the
distribution or in the fragmentation part. 
The asymmetries in Table~\ref{tabel2} are due to
\begin{eqnarray*}
\mbox{eo:}&\quad
&\mbox{polarized target}
\ \stackrel{\mbox{\scriptsize T-even}}{\longrightarrow}
\ \mbox{quark}^\uparrow 
\ \stackrel{\mbox{\scriptsize T-odd}}{\longrightarrow}
\ \mbox{unpolarized\ hadron},
\\
\mbox{oe:}&&
\mbox{target}^\uparrow 
\ \stackrel{\mbox{\scriptsize T-odd}}{\longrightarrow}
\ \mbox{unpolarized\ quark}
\ \stackrel{\mbox{\scriptsize T-even}}{\longrightarrow}
\ \mbox{unpolarized\ hadron},
\end{eqnarray*}
those in Table~\ref{tabel3} are due to
\begin{eqnarray*}
\mbox{eo:}&\quad&
\mbox{unpolarized\ target}
\ \stackrel{\mbox{\scriptsize T-even}}{\longrightarrow}
\ \mbox{unpolarized\ quark}
\ \stackrel{\mbox{\scriptsize T-odd}}{\longrightarrow}
\ \mbox{hadron}^\uparrow,
\\
\mbox{oe:}&&
\mbox{unpolarized\ target} 
\ \stackrel{\mbox{\scriptsize T-odd}}{\longrightarrow}
\ \mbox{quark}^\uparrow 
\ \stackrel{\mbox{\scriptsize T-even}}{\longrightarrow}
\ \mbox{polarized hadron},
\end{eqnarray*}
where the up-arrow denotes transversely polarized quarks or hadrons.

Next we turn to double spin asymmetries. These contain either both
T-even or both T-odd distribution and fragmentation functions. In
Table~\ref{tabel4} we have repeated only those even-even 
combinations from Ref.~\cite{mt96} that do not involve azimuthal 
angles in combination with azimuthal spin angle of the produced hadron. 
There exists only one odd-odd asymmetry at leading order.
This is the last entry in Table~\ref{tabel4}. 
\begin{table}[htb]
\caption{\label{tabel4}\protect
Some leading order even-even double spin asymmetries in leptoproduction
and the only leading order odd-odd double spin asymmetry.}
\begin{tabular}{cclc}
\\[-2 mm]
$ABC$ & $W$ 
& $\left<W\right>_{ABC} \cdot \left[4\pi\,\alpha^2\,s/Q^4\right]^{-1} $
& T 
\\[2mm] \hline 
\\[-2 mm]
LLO &
1 & 
$\lambda_e\,\lambda\,y\,(1-\frac{1}{2}\,y) \sum_{a,\bar a} e_a^2
\,\xbj\,g_1^a(\xbj) D_1^a(\zh)$
& ee 
\\[1mm]
LTO &
$(Q_\st/M)\,\cos(\phi_h^\ell-\phi_S^\ell)$ & 
$\lambda_e\,\vert \bm S_\st\vert\,y\,(1-\frac{1}{2}\,y) \sum_{a,\bar a} e_a^2
\,\xbj\,g_{1T}^{(1)a}(\xbj) D_1^{a}(\zh)$
& ee 
\\[1mm]
LOL &
1 & 
$\lambda_e\lambda_h\,y\left(1-\frac{1}{2}\,y\right) \sum_{a,\bar a} e_a^2
\,\xbj\,f_1^a(\xbj) G_1^a(\zh)$
& ee 
\\[1mm]
OLL &
1 & 
$\lambda\lambda_h
\,\left( 1-y + \frac{1}{2}y^2\right) \sum_{a,\bar a} e_a^2
\,\xbj\,g_1^a(\xbj) G_1^a(\zh)$
& ee 
\\[1mm]
OTT &
$\cos (\phi_S^\ell + \phi_{S_h}^\ell)$ &
$-\vert \bm S_{\st}\vert\,\vert \bm S_{h\st}\vert
\,\frac{1}{2}\,(1-y) \sum_{a,\bar a} e_a^2
\,\xbj\,h_1^a(\xbj)\,H_1^a(\zh)$ 
& ee 
\\[1mm] \hline 
\\[-2 mm]
OTT &
$(Q_\st^2/MM_h)\,\cos(\phi_S^\ell-\phi_{S_h}^\ell)$  & 
$\vert \bm S_{\st}\vert\,\vert \bm S_{h\st}\vert
\,\left( 1-y + \frac{1}{2}\,y^2\right) \sum_{a,\bar a} e_a^2
\,\xbj\,f_{1T}^{\perp (1) a}(\xbj) D_{1T}^{\perp (1) a}(\zh)$
& oo
\end{tabular}
\end{table}
The even-even asymmetries include LLO, LOL and OLL asymmetries where compared 
with the $\left< 1 \right>_{OOO}$ result pairs of unpolarized particles 
are replaced by longitudinally polarized particles. The LTO asymmetry
is the even-even equivalent of the odd-even OTO single spin asymmetry
in Table~\ref{tabel2} and is probably the easiest way to obtain the 
function $g_{1T}^{\perp (1)}$, e.g. in leptoproduction of pions
$\vec e p^\uparrow \rightarrow e \pi^+ X$~\cite{km96}.   
The even-even OTT asymmetry has been suggested by Artru as the way to 
obtain the transverse spin distribution $h_1$ in the proton 
via e.g.\ $ep^\uparrow \rightarrow e\Lambda^\uparrow X$~\cite{artru}.
In the same process the odd-odd asymmetry can be investigated.
While in the even-even Artru asymmetry
a (longitudinal) virtual photon scatters off a transversely polarized quark,
one has in the odd-odd asymmetry a (transverse) virtual photon scattering
off an unpolarized quark.

It is useful at this point to mention that the asymmetries with a 
fragmentation function $D_1$ can also be obtained by looking at the
asymmetry in jet-production. In that case one needs
the fragmentation of a quark into a quark, which (at tree-level) is
given by $D_1(z) = \delta(1-z)$. One then can
perform the $z_h$-integration. Thus the azimuthal asymmetry
$\cos(\phi_{jet}^\ell - \phi_S^\ell)$ is a way to probe
$g_{1T}$. The first result in Table~\ref{tabel4} even survives 
after full integration over the final states, giving the ordinary double 
spin asymmetry in inclusive leptoproduction 
in terms of the polarized quark distribution function $g_1$.

Finally, for completeness, we give in Table~\ref{tabel5}
the two possible leading order triple spin asymmetries with a T-odd
distribution function for 
{\em polarized leptons scattering off a transversely polarized target} 
leading to a spin asymmetry in the final state.
The first asymmetry is the analogue of the OTO single spin asymmetry
in Table~\ref{tabel2} with the unpolarized particles replaced by
longitudinally polarized particles.
\begin{table}[htb]
\caption{\label{tabel5}\protect
The leading order triple spin asymmetries with a T-odd
distribution functions in leptoproduction.}
\begin{tabular}{cclc}
\\[-2 mm]
$ABC$ & $W$ 
& $\left<W\right>_{ABC} \cdot \left[4\pi\,\alpha^2\,s/Q^4\right]^{-1} $
& T
\\[2mm] \hline 
\\[-2 mm]
LTL &
$(Q_\st/M)\,\sin(\phi_h^\ell-\phi_S^\ell)$  & 
$\lambda_e\,\vert \bm S_\st\vert\,\lambda_h
\,y\,(1-\frac{1}{2}\,y) \sum_{a,\bar a} e_a^2
\,\xbj\,f_{1T}^{\perp (1) a}(\xbj) G_{1}^{a}(\zh)$
& oe
\\[1mm]
LTT &
$(Q_\st^2/2MM_h)\,\sin(\phi_S^\ell-\phi_{S_h}^\ell)$  & 
$\lambda_e\,\vert \bm S_\st\vert\,\vert \bm S_{h\st}\vert
\,y\,(1-\frac{1}{2}\,y) \sum_{a,\bar a} e_a^2
\,\xbj\,f_{1T}^{\perp (1) a}(\xbj) G_{1T}^{(1) a}(\zh)$
& oe
\end{tabular}
\end{table}

At leading order the T-odd distribution functions only appear in
azimuthal asymmetries. At subleading ($1/Q$) order T-odd distribution 
functions appear also in
the simple $\bm q_T$-integrated cross sections, i.e.\ the ones that do
not involve powers of $Q_T$ and azimuthal angle $\phi_h^\ell$
in the weight function, but at most the
azimuthal spin angles ($\phi_S^\ell$ or $\phi_{S_h}^\ell$). 
In that case all that is needed are the
soft parts integrated over transverse momenta. At subleading order,
however, one needs to include parts proportional to $M/P^+$ in $\Phi$ 
and parts proportional to $M_h/P_h^-$ in $\Delta$. In the cross
sections these factors give rise to a suppression factor $1/Q$. 
The quantity needed in the calculation is
\bea
\Phi(x) &\equiv& \int d^2\bm p_\st\,\Phi(x,\bm p _\st) =
\frac{1}{2}\,\Biggl\{ f_1\,\slash \slash n_+ 
+\lambda\,g_1\, \gamma_5\slash n_+
+h_1\,\frac{[\slash S_\st,\slash n_+]\gamma_5}{2}
\Biggr\}
\nonumber \\ & & 
+\frac{M}{2P^+}\,\Biggl\{
f_T\,\epsilon_\st^{\rho\sigma}S_{\st\rho}\gamma_\sigma
+ e\,{\bm 1} - i\,\lambda\,e_L\,\gamma_5 +g_T\, \gamma_5\slash S_\st
+\lambda \,h_L\,\frac{[\slash n_+, \slash n_-]\gamma_5}{2}
+i\,h\,\frac{[\slash n_+, \slash n_-]}{2}
\Biggr\},
\eea
in terms of distribution functions with arguments $f_1$ = $f_1(x)$ etc.
All the leading twist (twist-two) functions are T-even.
Of the twist-three functions (multiplying $(M/P^+)$) the 
functions $f_T$, $e_L$ and $h$ are T-odd ones. The function $f_T$
is also discussed in Ref.~\cite{teryaev} ($f_T \propto c_V$).
Noteworthy is the relation between some of the $\bm p_\st$-integrated
twist-three functions and $\bm p_\st^2/M^2$ moments of leading 
$\bm p_\st$-dependent distribution functions~\cite{mt96},
\bea
&&g_T(x) = g_1(x) + \frac{d}{dx}\,g_{1T}^{(1)},
\\
&&h_L(x) = h_1(x) - \frac{d}{dx}\,h_{1L}^{\perp(1)},
\\
&&f_T(x) = - \frac{d}{dx}\,f_{1T}^{\perp(1)},
\\
&&h(x) = - \frac{d}{dx}\,h_{1}^{\perp(1)}.
\eea
For the first case this relation appears in a slightly different
form (using quark-quark-gluon correlation functions) in Ref.~\cite{bkl84}.
For the fragmentation part one needs at order
$1/Q$ the quantity
\bea
\Delta(z) &\equiv& z^2\int d^2 \bm k_\st\ \Delta(z,\bm k_\st) =
\frac{1}{2}\,\Biggl\{
D_1\,\slash \slash n_-
+\lambda_h\,G_1\, \gamma_5\slash n_-
+H_1\,\frac{[\slash S_{h\st},\slash n_-]\gamma_5}{2}
\Biggr\}
\nonumber \\ & & 
+\frac{M_h}{2P_h^-}\,\Biggl\{
D_T\,\epsilon_\st^{\rho\sigma}\gamma_\rho S_{h\st\sigma}
+ E\,{\bm 1} - i\,\lambda_h\,E_L\,\gamma_5
+G_T\, \gamma_5\slash S_{h\st}
+\lambda_h \,H_L\,\frac{[\slash n_-, \slash n_+]\gamma_5}{2}
+i\,H\,\frac{[\slash n_-, \slash n_+]}{2}
\Biggr\},
\eea
in terms of fragmentation functions with arguments $D_1$ = $D_1(z)$ 
= $\int d^2\bm k_\st^\prime \,D_1(z,\bm k_\st^2)$, etc.
For spin zero particles (e.g. pions) only the twist-two function $D_1$ and
the twist-three functions $E$ and $H$ appear, the latter one being T-odd.
Some of the $\bm k_\st$-integrated twist-three functions were already 
mentioned in 
Ref.~\cite{ji94} ($E \propto \hat e_1$ and $H \propto \hat e_{\bar 1}$).
The function $D_T$ was also discussed in Refs~\cite{lu,bjm97}.
The relations between $\bm k_\st$-integrated twist-three functions and 
$\bm k_\st^2/2M_h^2$ moments of $\bm k_\st$-dependent fragmentation 
functions are
\bea
&&\frac{G_T(z)}{z} = \frac{G_1(z)}{z} - z^2\,\frac{d}{dz}\,\left[
\frac{G_{1T}^{(1)}(z)}{z}\right],
\\
&&\frac{H_L(z)}{z} = \frac{H_1(z)}{z} + z^2\,\frac{d}{dz}\,\left[
\frac{H_{1L}^{\perp(1)}(z)}{z}\right],
\\
&&\frac{D_T(z)}{z} = z^2\,\frac{d}{dz}\,\left[
\frac{D_{1T}^{\perp(1)}(z)}{z}\right],
\\
&&\frac{H(z)}{z} = z^2\,\frac{d}{dz}\,\left[
\frac{H_{1}^{\perp(1)}(z)}{z}\right].
\eea
In the calculation of the hadron tensor not only the quark-quark correlation
functions in $\Phi$ and $\Delta$ need to be considered, but as well
quark-quark-gluon correlation functions, which contain transverse gluon
fields. With the help of the equations of motion, however, the subleading
contribution in the cross sections can be expressed in terms of the 
twist-three quark-quark correlation functions~\cite{jj}. 
The fragmentation functions appear in specific combinations
\bea
&&
\frac{\tilde D_T(z)}{z} = 
\frac{D_T(z)}{z} 
+ D_{1T}^{\perp (1)}(z) 
\\
&&
\frac{\tilde E(z)}{z} = 
\frac{E(z)}{z} - \frac{m}{M_h}\,D_1(z)
\\
&&
\frac{\tilde E_L(z)}{z} = 
\frac{E_L(z)}{z} 
\\
&&
\frac{\tilde G_T(z)}{z} = 
\frac{G_T(z)}{z} 
- \frac{m}{M_h}\,H_1(z) - G_{1T}^{(1)}
\\
&&
\frac{\tilde H_L(z)}{z} = 
\frac{H_L(z)}{z} 
- \frac{m}{M_h}\,G_1(z) + 2\,H_{1L}^{\perp (1)}(z)
\\
&&
\frac{\tilde H(z)}{z} = 
\frac{H(z)}{z} + 2\,H_1^{\perp (1)}(z)
\eea
which are the truely interaction-dependent parts of the twist-three 
functions~\cite{mt96}. 
The results for the asymmetries at subleading order are given 
in Table~\ref{tabel6}.
\begin{table}[htb]
\caption{\label{tabel6}\protect
The subleading order spin asymmetries in leptoproduction.}
\begin{tabular}{cllc}
\\[-2 mm]
$ABC$ & $W$ 
& $\left<W\right>_{ABC} \cdot \left[4\pi\,\alpha^2\,s/Q^4\right]^{-1} $
& T
\\[2mm] \hline 
\\[-2 mm]
LTO &
$\cos \phi_S^\ell$ &
$-\lambda_e\,\vert\bm S_\st\vert
\,y\,\sqrt{1-y} \ \sum_{a,\bar a} e_a^2
\bigl[\frac{M\xbj^2}{Q}\,g_T^a(\xbj) D_1^a(\zh)
+\frac{M_h\xbj}{\zh Q}\,h_1^a(\xbj) \tilde E^a(\zh)\bigr]$
& ee/ee
\\[3mm]
OTO &
$\sin \phi_S^\ell$ &
$\vert \bm S_{\st}\vert
\,(2-y)\sqrt{1-y} \ \sum_{a,\bar a} e_a^2
\bigl[\frac{M_h\xbj}{\zh Q}\,h_1^a(\xbj)\tilde H^a(\zh)
-\frac{M\xbj^2}{Q}\,f_T^a(\xbj)\,D_1^a(\zh)\bigr]$
& eo/oe
\\[3mm]
LOT &
$\cos \phi_{S_h}^\ell$ &
$-\lambda_e\,\vert\bm S_{h\st}\vert
\,y\,\sqrt{1-y} \ \sum_{a,\bar a} e_a^2
\bigl[\frac{M\xbj^2}{Q}\,e^a(\xbj) H_1^a(\zh)
+\frac{M_h\xbj}{\zh Q}\,f_1^a(\xbj) \tilde G^a(\zh)\bigr]$
& ee/ee
\\[3mm]
OOT &
$\sin \phi_{S_h}^\ell$ &
$\vert \bm S_{h\st}\vert \,(2-y)\sqrt{1-y} \ \sum_{a,\bar a} e_a^2
\,\bigl[\frac{M_h\xbj}{\zh Q}\,f_1^a(\xbj)\tilde D_T^a(\zh)
-\frac{M\xbj^2}{Q}\,h^a(\xbj)\,H_1^a(\zh)\bigr]$
& eo/oe
\\[3mm]
OTL &
$\cos \phi_S^\ell$ &
$-\vert\bm S_\st\vert\,\lambda_h
\,(2-y)\sqrt{1-y} \ \sum_{a,\bar a} e_a^2
\bigl[\frac{M\xbj^2}{Q}\,g_T^a(\xbj) \,G_1^a(\zh)
+\frac{M_h\xbj}{\zh Q}\,h_1^a(\xbj) \tilde H_L^a(\zh) \bigr]$
& ee/ee
\\[3mm]
LTL &
$\sin \phi_S^\ell$ &
$-\lambda_e\,\vert\bm S_{\st}\vert\,\lambda_h
\,y\,\sqrt{1-y} \ \sum_{a,\bar a} e_a^2
\bigl[\frac{M_h\xbj}{\zh Q}\,h_1^a(\xbj) \tilde E_L(\zh)
+\frac{M\xbj^2}{Q}\,f_T^a(\xbj) \,G_1^a(\zh)\bigr]$
& eo/oe
\\[3mm]
OLT &
$\cos \phi_{S_h}^\ell$ &
$-\lambda\,\vert\bm S_{h\st}\vert
\,(2-y)\sqrt{1-y} \ \sum_{a,\bar a} e_a^2
\bigl[ \frac{M\xbj^2}{Q}\,h_L^a(\xbj) \,H_1^a(\zh)
+\frac{M_h\xbj}{\zh Q}\,g_1^a(\xbj) \tilde G_T^a(\zh)\bigr]$
& ee/ee
\\[3mm]
LLT &
$\sin \phi_{S_h}^\ell$ &
$\lambda_e\lambda\,\vert\bm S_{h\st}\vert
\,y\,\sqrt{1-y} \ \sum_{a,\bar a} e_a^2
\bigl[\frac{M_h\xbj}{\zh Q}\,g_1^a(\xbj) \tilde D_T(\zh)
+\frac{M\xbj^2}{Q}\,e_L^a(\xbj) \,H_1^a(\zh)\bigr]$
& eo/oe
\end{tabular}
\end{table}

There are two asymmetries for leptoproduction of spin zero particles
(e.g.\ pions or kaons), the first two entries in Table~\ref{tabel6}.
The first one is an LTO asymmetry which contains, like all asymmetries 
in the Table, two terms. The first one is the one which survives in
inclusive leptoproduction when one sums over all final states. The 
presence of the second term shows that using production of specific
hadrons, e.g.\ strange ones to tag strange quarks cannot be used to
disentangle different flavor contributions $g_T^a$. 

At order $1/Q$ and integrating over the transverse momentum of the
produced hadrons, there exist two single spin asymmetries, an OTO and
an OOT asymmetry. They involve odd-even and even-odd combinations
of distribution and fragmentation functions. Other such combinations
lead to triple spin asymmetries. The ordinary double-spin
asymmetries involve only even-even combinations. Since for the
transverse momentum averaged correlation functions 
the T-odd functions only appear at the twist-three level, 
any odd-odd combination is of order $1/Q^2$. 

In conclusion, we have presented leading order azimuthal asymmetries
involving the azimuthal angles of the transverse momentum of the 
produced hadron or of the spin vectors of any of the hadrons involved,
i.e.\ the target hadron or the produced hadron. Furthermore, for the
transverse momentum integrated case we have given the results up to
order $1/Q$. One of the reasons are the relations between the
twist-three functions relevant at subleading order and the transverse
momentum dependent functions, which exist for both distribution and
fragmentation functions. Several isolated cases have been pointed out
before, but we have presented a systematic overview including in
particular a number of new asymmetries that could facilitate experimental
searches for the recently much debated T-odd fragmentation functions.

We would like to acknowledge useful discussions with M. Anselmino (Torino),
A. Drago (Ferrara) and R. Jakob (NIKHEF/Pavia).
This work was supported by the Foundation for Fundamental Research on
Matter (FOM) and the Dutch Organization for Scientific Research (NWO).

\end{document}